# Picosecond Supercontinuum Generation in All-Normal Dispersion Optical Fibers Enabled by Polarization Instabilities

**R. MOREL[1], A. KUDLINSKI[2], O. VANVINCQ[2], L. EMONIN[1], G. FANJOUX[1], J. M. DUDLEY[1], T. SYLVESTRE[1,*]**

[1] *Université Marie et Louis Pasteur, CNRS, Institut FEMTO-ST, Besançon, France*
[2] *Université de Lille, CNRS, PhLAM-Physique des Lasers Atomes et Molécules, F-59000, Lille, France*
*[thibaut.sylvestre@umlp.fr](mailto:thibaut.sylvestre@umlp.fr)

**Supercontinuum generation in all-normal-dispersion optical fibers has so far been predominantly explored under femtosecond pumping conditions. Here, we demonstrate that efficient and broadband supercontinuum generation can also be achieved in the long picosecond regime by pumping a highly birefringent all-normal-dispersion silica-based photonic crystal fiber at 1064 nm. The observed spectral broadening results from the combined action of polarization modulation instability and cascaded Raman scattering, enabling octave-spanning spectra extending from 600 nm to 1650 nm. These results establish a distinct operating regime for supercontinuum generation and open new perspectives for robust, high-power broadband fiber sources.**

Supercontinuum (SC) generation in all-normal-dispersion (ANDi) optical fibers has attracted significant attention over the past decade owing to its ability to deliver broadband, spectrally flat light with high coherence, low noise, and excellent long-term stability [1-4]. These characteristics are highly attractive for demanding applications such as optical coherence tomography (OCT), coherent beam combining, nonlinear microscopy, ultrafast pulse synthesis, and shot-noise-limited dual-comb spectroscopy [5-7]. Unlike soliton-based SC generation in the anomalous-dispersion regime, which is intrinsically sensitive to noise amplification through modulation instability, soliton interactions, and stimulated Raman scattering (SRS), femtosecond-driven ANDi SC relies primarily on coherent self-phase modulation (SPM) and optical wave breaking (OWB). This leads to deterministic spectral broadening with superior coherence and reduced noise compared with soliton-based SC generation [1-5].

To date, ANDi SC generation has been almost exclusively investigated under femtosecond pumping conditions, where the interplay between SPM and OWB governs the nonlinear dynamics and enables flat, symmetric coherent SC spectra [1]. In contrast, the picosecond pumping regime remains largely unexplored in ANDi fibers, despite its strong practical relevance. Picosecond lasers offer higher pulse energies, improved robustness, and superior scalability compared to femtosecond systems, making them particularly attractive for high-power broadband light sources. However, extending ANDi SC generation into the picosecond regime presents fundamental challenges. Longer pulse durations suppress OWB while enhancing the influence of SRS and nonlinear polarization instabilities, which are commonly associated with degraded coherence and increased noise [8]. In particular, polarization modulation instability (PMI) can play a significant role in picosecond-pumped ANDi fibers, especially in weakly birefringent structures, leading to polarization-dependent spectral distortions and reduced stability [8]. Although polarization-maintaining (PM) ANDi fibers have been developed to mitigate these effects, many existing designs rely on stress-applying elements or complex microstructured claddings, which increase fabrication complexity and limit design flexibility [9]. Moreover, the combined impact of PMI and SRS on SC generation in the picosecond regime within PM-ANDi fibers has not yet been systematically investigated.

In this Letter, we experimentally demonstrate SC generation in the long-pulse regime using a PM-ANDi photonic crystal fiber (PCF). A highly birefringent ANDi PCF is pumped at 1064 nm with 42-ps pulses from a mode-locked laser operating at a repetition rate of 200 kHz. We show that the SC dynamics are strongly polarization dependent, with efficient spectral broadening occurring only when the fiber is pumped along its slow axis. In this configuration, PMI is activated and, in combination with cascaded SRS, enables efficient spectral expansion. As a result, a wide and flat SC spanning from 600 nm to 1650 nm is generated, representing the maximum bandwidth achieved in our experiments. These results establish a previously unexplored operating regime for SC generation, bridging femtosecond-driven coherent broadening and picosecond nonlinear dynamics. The experimental observations are compared with numerical simulations based on the coupled generalized nonlinear Schrödinger equations, together with analytical phase-matching conditions used to predict the PMI sidebands.

The PCF investigated here was recently applied to femtosecond-pumping SC generation in Ref. [10]; its main properties are summarized in Fig. 1. The scanning electron microscope (SEM) image of the cross-section is displayed in the inset of Fig. 1(b). The microstructure has an average hole-to-hole spacing of 1.42 µm and average hole diameter of 0.55 µm, except for two slightly bigger holes (0.64 µm) located across the core (Fig. 1(d)). Simulations of the fiber properties were performed with a finite-element method (FEM) mode solver based on the SEM image. For clarity, we present only the results along the slow axis of the fiber, as those along the fast axis are nearly identical. As shown in Fig. 1(a), the group-velocity dispersion (GVD) is entirely normal, with a maximum of -16 ps·nm⁻¹·km⁻¹ at 1000 nm, in good agreement with measurements (red curve), obtained using the method of Ref. [11]. Confinement losses remain below 100 dB/km up to 1600 nm (Fig. 1(b)). The two enlarged holes induce a phase birefringence $\delta n = 2 \times 10^{-4}$ at 1000 nm and a group birefringence $\delta n_G$ that is negative at 1000 nm and changes sign near 1300 nm, again in good agreement with measurements (in red in Fig. 1(c)), using the method from Ref. [12]. Note that we use the convention that the fast axis (FA) is the axis with the lowest effective index and slow axis (SA) the axis with the highest effective. Figure 1(d) shows the mode profile recorded with a CCD camera, together with the intensity profiles along two orthogonal directions. The average mode-field diameter is 2.14 µm with an ellipticity of 95.4%. The PCF operates experimentally in the single-mode regime, as expected from its air-fill fraction of 0.39.

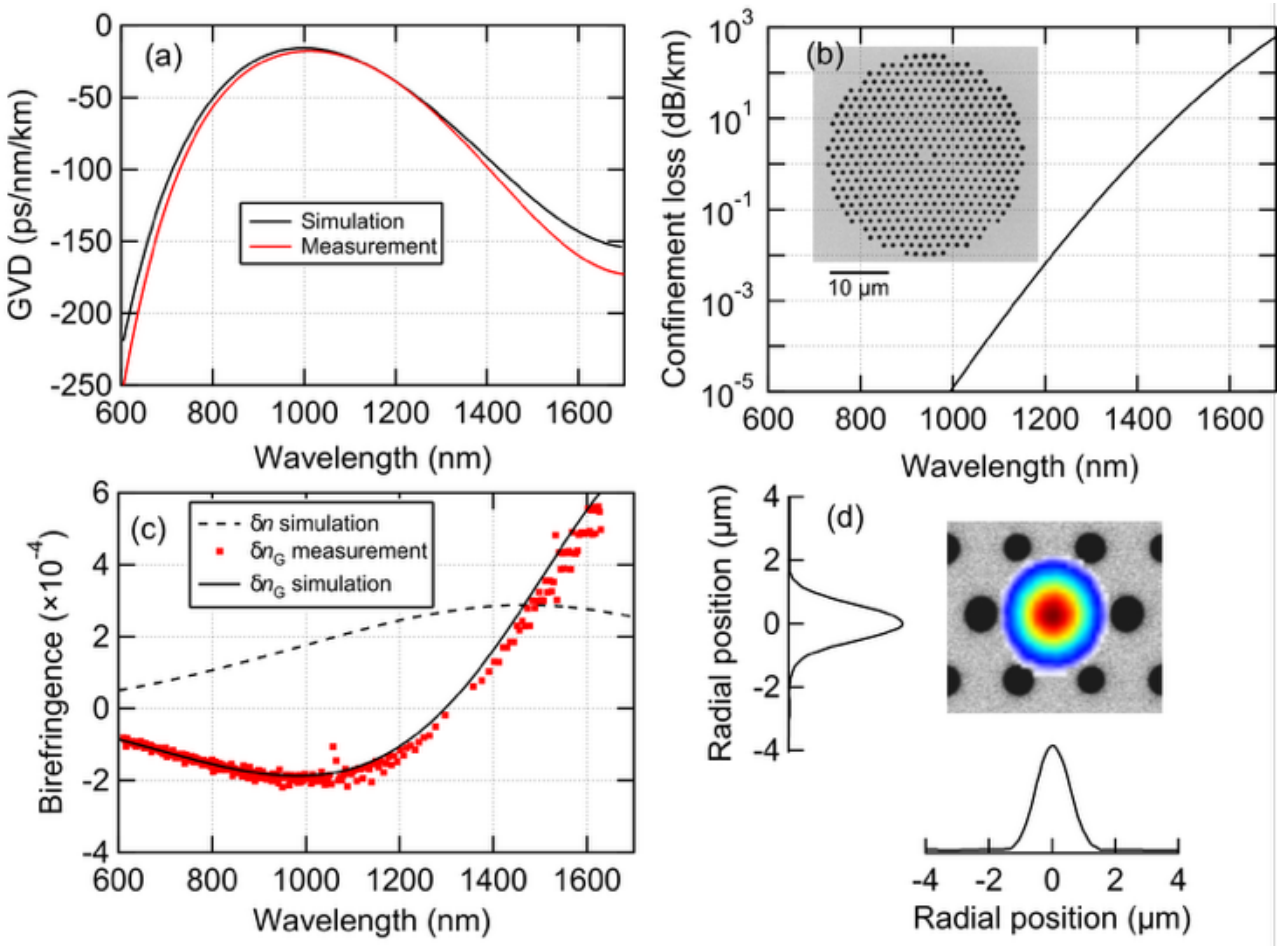

Fig. 1. Properties of the PM-ANDi PCF. (a) Measured (red) and simulated (black) GVD along the slow axis. (b) Simulated confinement loss of the slow axis (inset: SEM image of the PCF). (c) Simulated phase birefringence $\delta n$ (dashed line) and group birefringence $\delta n_G$ (solid black line), with its measurement in red dots. (d) Close-up of the PCF core with the measured mode profile superimposed and measured intensity profiles along both polarization axes.

Experimentally, the fiber was pumped using a high-power picosecond laser (EKSPLA Atlantic 5) operating at 1064 nm, delivering 42 ps pulses at a repetition rate of 200 kHz and providing up to 4 W of average power. The input polarization was precisely controlled with a half-wave plate, and the beam was coupled into the fiber using an aspherical lens, achieving efficient mode matching and a coupling efficiency of approximately 60%. The output spectra were measured using two optical spectrum analyzers (Yokogawa AQ6373 and AQ6374), covering the spectral ranges 350–1200 nm and 600–1700 nm, respectively. The output power was continuously monitored with a calibrated photodiode. In addition, a polarizer placed at the fiber output enabled detailed analysis of the polarization properties and measurement of the polarization extinction ratio (PER). Fiber lengths between 0.35 m and 10 m were used in experiments.

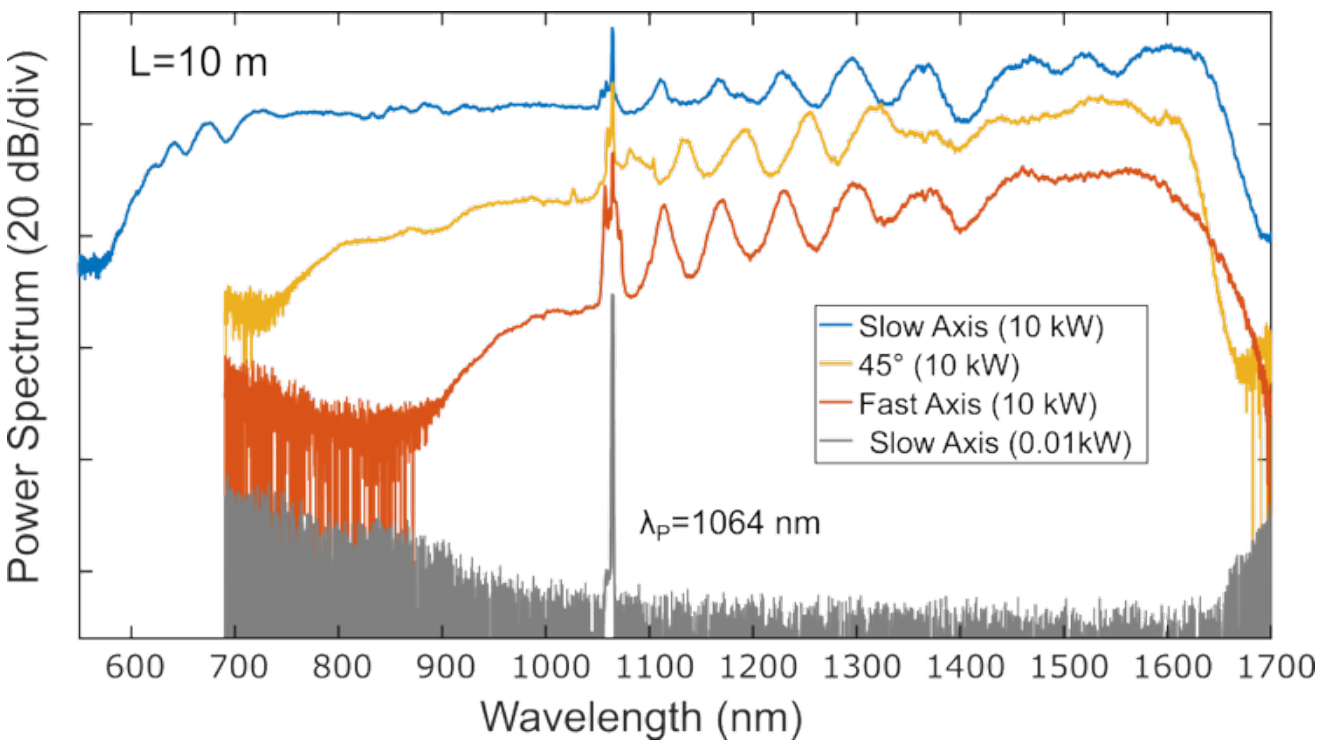

Fig. 2. Experimental spectra measured at the output of a 10 m PM-ANDi photonic crystal fiber for different input polarization states: slow axis (red), fast axis (blue), and 45° off-axis (yellow), at an input peak power of 10 kW. The gray curve shows, for comparison, the linear regime with the laser peak at 1064 nm for a peak power of 0.01 kW. The spectra are plotted in arbitrary units and offset by 10 dB for clarity.

We first investigated SC generation in the longer 10 m fiber section for different input polarization states. Figure 2 shows the experimental output spectra, highlighting that the broadest SC is achieved when pumping along the slow axis (blue trace), with a bandwidth spanning 600 nm to 1650 nm at –20 dB from the peak. Remarkably, this spectral span exceeds that typically achieved in the femtosecond regime (650–1350 nm), despite the strong confinement losses beyond 1350 nm. Figure 2 also clearly demonstrates the strong dependence of the SC bandwidth on the input polarization: approximately 1000 nm for slow-axis pumping (blue), 613 nm for fast-axis pumping (red), and about 800 nm for 45° off-axis excitation (yellow). While the Raman cascade is essentially similar for both principal axes, thereby defining the long-wavelength edge of the SC near 1650 nm, the main differences arise on the anti-Stokes side of the spectrum. In particular, for fast-axis pumping, the SC does not extend below 900 nm. In contrast, under 45° off-axis excitation, the SC broadens down to 800 nm. In this configuration, the Raman cascade is spectrally shifted due to cross-phase modulation instability (XPMI), which generates a pronounced four-wave mixing (FWM) Stokes sideband approximately 7 THz from the pump [13,14]. This sideband effectively acts as a secondary pump, inducing a 7 THz shift of the Raman cascade and significantly reshaping the overall SC dynamics.

To gain further insight on SC mechanisms, we investigated shorter fiber segments of 35 cm and 75 cm, while pumping along the fiber slow axis. The results are shown in Figs. 3(a–b) for two increasing peak powers. For a peak power of 9 kW (green spectra), far-detuned anti-Stokes (AS) and Stokes (S) sidebands appear at 865.4 nm and 1380.7 nm (~64.7 THz far-detuned from the pump frequency), cross-polarized along the fast axis with respect to the pump, while Raman peaks ($R_1$ and $R_{-1}$) remain weak. These sidebands arises from PMI and satisfies the phase-matching condition and parametric gain, associated with vector degenerate four-wave mixing (FWM), $2\omega_P = \omega_S + \omega_{AS}$, as described by the following equations [15],

$$\Delta\beta = \beta_2\delta\omega^2 + \frac{\beta_4}{12}\delta\omega^4 - \frac{2\omega_P\delta n}{c} - \frac{2}{3}\gamma P = 0 \qquad (1)$$

$$g = Re\left(\sqrt{\left(\frac{\gamma P}{3}\right)^2 - \left(\frac{\Delta\beta}{2}\right)^2}\right) \qquad (2)$$

where $\beta_2$ and $\beta_4$ are the second and fourth-order dispersions at 1064 nm for the fast axis. $\delta\omega = \omega_P - \omega_S$ is the frequency shift between the pump frequency $\omega_P$ and the PMI sidebands. $\gamma$ the nonlinear coefficient and $c$ the speed of light in vacuum. Figure 4 shows the predicted PMI peak wavelengths and the corresponding parametric gain $g$, calculated from the above equations, as a function of the input pump wavelength, using the parameters from Fig. 1. At 1064 nm, the model predicts PMI gain at 866 nm and 1380 nm, corresponding to a frequency shift of 64.5 THz. This value is in good agreement with the experimental shift of 64.7 THz observed in Fig. 3(a). Note that such large frequency shift has already been reported in weaky birefringent PCFs [15]. Interestingly, Fig. 4 shows that the maximum frequency shift (70 THz) does not occur near the dispersion maximum (1000 nm) but at 950 nm. This behavior is due to the non-negligible contribution of fourth-order dispersion arising from the parabolic dispersion profile, which has often been neglected in previous studies on PMI [13].

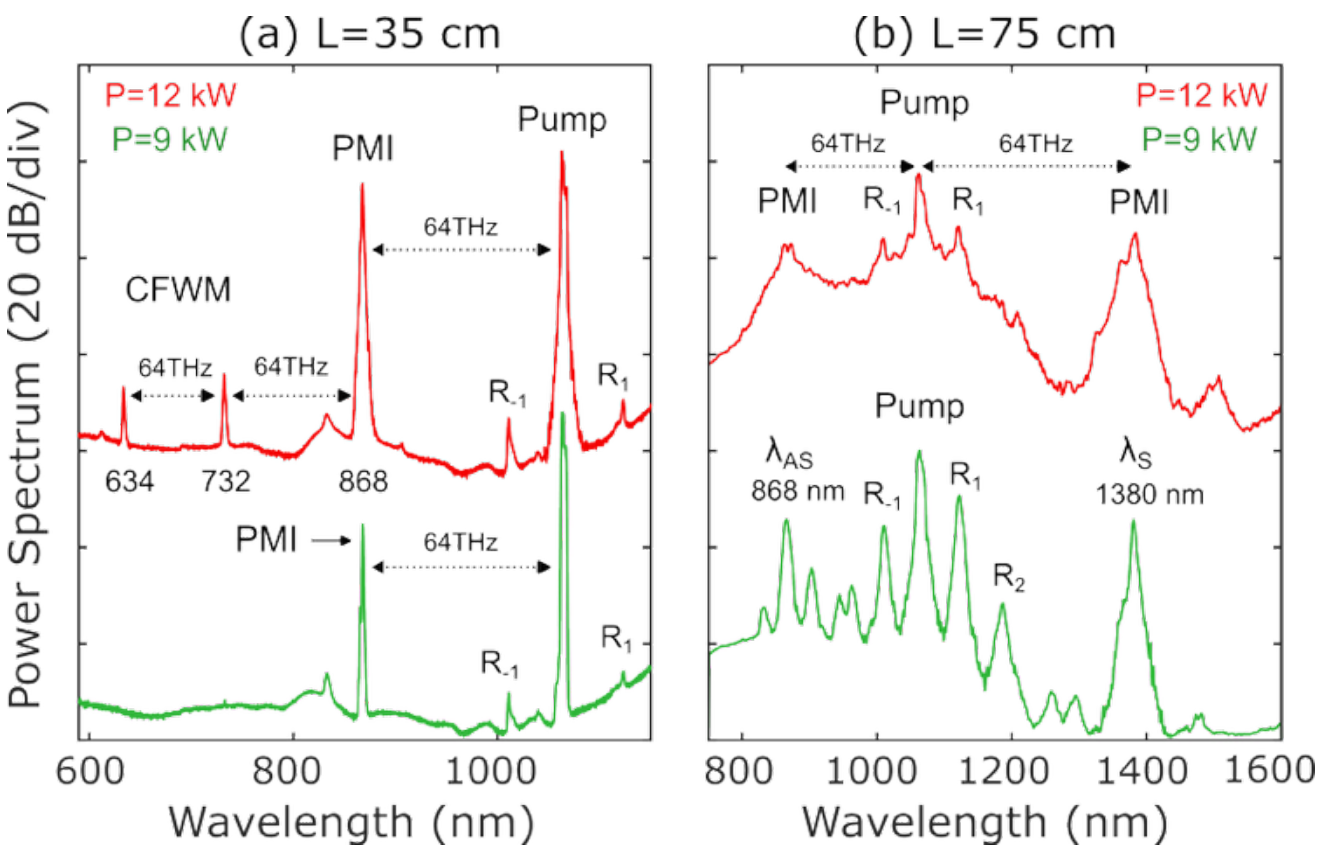

Fig. 3. Experimental output spectra for fibers of two lengths, (a) L = 35 cm and (b) L = 75 cm, measured at two input peak powers (9 kW in green, 12 kW in red) with slow-axis (SA) pumping, highlighting PMI and Raman sidebands contributing to SC generation.

In addition, Fig. 3(a) shows the generation of two higher-order anti-Stokes sidebands at higher peak power (12 kW, red spectrum), detuned by up to 192 THz from the pump. These sidebands originate from cascaded four-wave mixing (CFWM) with alternating cross-polarizations. Although similar behavior has been reported in isotropic fibers and fiber cavities [16,17], such large frequency shifts, particularly the third-order peak at 192 THz from the pump (634 nm), have not been reported previously. Figure 3(b) shows that increasing the fiber length to 75 cm further enhances the spectral broadening of both the PMI sidebands and the Raman components (green spectrum), ultimately leading to SC generation at a peak power of 12 kW (red spectrum). The PMI Stokes sideband at 1380 nm is also observed in Fig. 3(b), in good agreement with the theoretical prediction shown in Fig. 4.

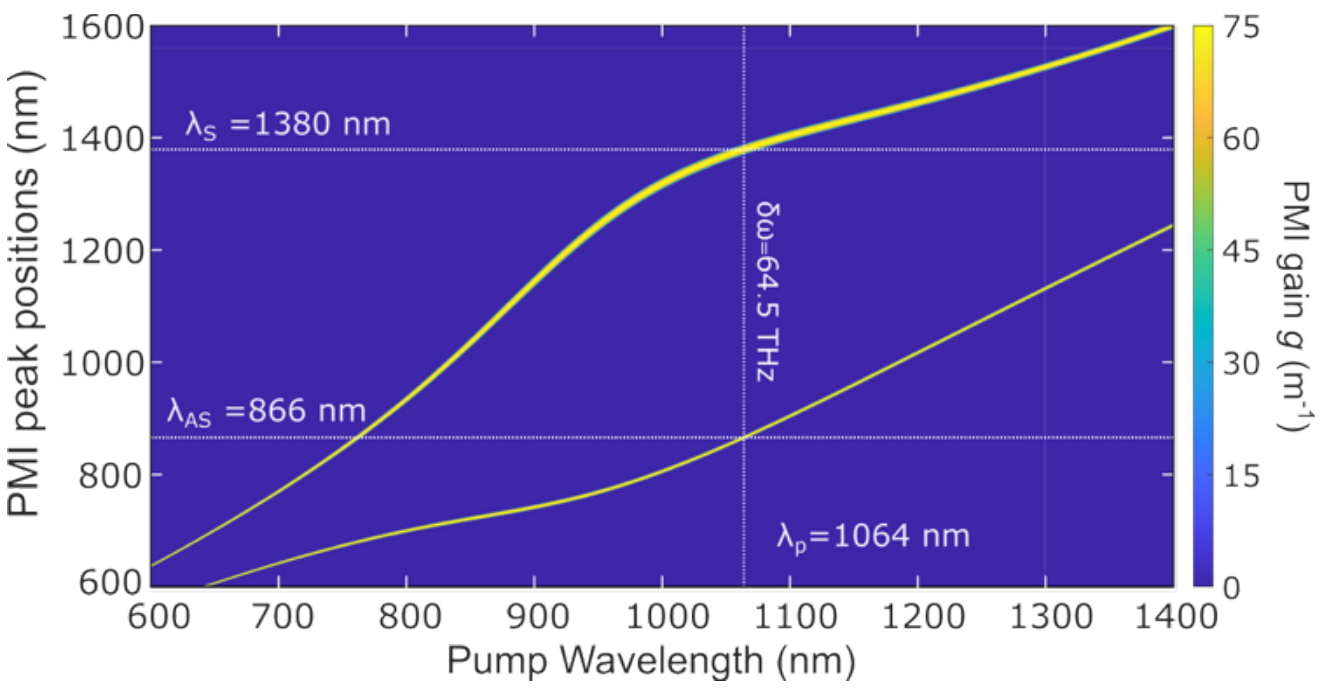

Fig. 4. PMI gain (Eq. 2) as a function of the pump wavelength. The vertical line indicates the pump wavelength at 1064 nm, while the two horizontal lines mark the PMI wavelengths. The parameters used in the calculation for 1064 nm are $\beta_2 = 1.1\times10^{-26}$ $s^2\cdot m^{-1}$, $\beta_4 = 3.7\times10^{-55}$ $s^4\cdot m^{-1}$, $\delta n = 2\times10^{-4}$, $\gamma = 30$ $W^{-1}\cdot km^{-1}$, and P = 14 kW.

From Figs. 3(a-b), the origin of the flat SC shown in Fig. 2 can be more clearly understood. The cascaded PMI process, together with its interplay with SRS, which becomes more pronounced for longer fiber lengths, as illustrated in Fig. 3(b), enables the generation of such a broad spectrum. Notably, the first-order PMI sideband generates its own Raman component in parallel with that initiated by the pump. To further elucidate the underlying dynamics, numerical simulations were carried out using two coupled generalized nonlinear Schrödinger equations (CGNLSE) (Eq. (3)), with fiber losses neglected owing to the short propagation lengths. The equations are written as follows[13,18]:

$$\frac{\partial A_{x,y}}{\partial z} + \beta_{1_{x,y}}\frac{\partial A_{x,y}}{\partial t} - i\sum_{k=2}^{+\infty}\left[\frac{(-i)^k\beta_k}{k!}\right]\frac{\partial^k A_{x,y}}{\partial t^k}$$
$$= i\gamma\left(1 + i\tau_s\frac{\partial}{\partial t}\right)\left\{(1 - f_R)\left[\left(\left|A_{x,y}\right|^2 + \frac{2}{3}\left|A_{y,x}\right|^2\right)A_{x,y} + \frac{1}{3}A_{x,y}^{*}A_{y,x}^{2}e^{-2\,i\,\Delta\beta_0\,z}\right] + f_R R_{x,y}(t)\right\} \qquad (3)$$

where $A_x$ and $A_y$ are the slowly varying amplitudes of the linearly polarized field components along the slow and fast axis, respectively. $\beta_{1_{x,y}}$ denote the corresponding propagation constants, while $\beta_k$ are the dispersion coefficients from in Fig. 1(a) assumed identical for the two axis. $\gamma$ is the nonlinear coefficient, and $\tau_s$ represents the self-steepening (shock) term. $f_r$ is the fractional Raman contribution in silica, and $\Delta\beta_0 = \beta_{0x} - \beta_{0y}$ is the phase detuning between the two axes. Finally, $R_{x,y}(t)$ is the Raman response of silica for each axis, as defined in [13].

The simulations were performed using 42 ps Gaussian pump pulses at 1064 nm, linearly polarized along the slow axis of the ANDi fiber. Quantum noise was included at the level of one photon per mode [19], and the results correspond to an average of ten SC spectra obtained with different noise seeds. The simulations used the same parameters as before, with a peak power of 10 kW and a fiber length of 10 m.

Figure 5(a) shows a direct comparison between the numerical (red) and experimental (red dashed) and SC spectra, while Fig. 5(b) presents a heatmap of the power spectral density (PSD) along the fiber, highlighting the dominant role of the strong first-order PMI sidebands in SC generation. These sidebands, appearing at 866 nm and 1380 nm, emerge from quantum noise at an early propagation stage. Only ~0.3 m of fiber is sufficient to reach the SC regime, characterized by a continuous spectrum rather than discrete bands. After an additional ~3 m, the SC reaches a quasi-steady state and its maximum bandwidth. Notably, the strong spectral broadening around 0.3 m in Fig. 5(b), affecting all peaks including the pump, also indicates temporal coherence degradation. As expected for SC dynamics dominated by noise-amplification processes such as SRS and PMI, combined with long pulse duration and fiber length, the resulting SC exhibits low coherence, as shown by the solid blue curve in Fig. 5(a), where we plot the mutual coherence function $g_{12}$ following the approach of Ref. [19]. Furthermore, the strong polarization instability during nonlinear propagation completely erases the PM properties of the fiber. Indeed, using a broadband polarizer (LPVIS050, Thorlabs) at the fiber output, we measured a polarization extinction ratio of only ~3 dB across the SC bandwidth. Nevertheless, this approach enables complete suppression of the residual pump and improves spectral flatness, achieving over 50 µW·nm⁻¹ across the SC. The PSD could be further increased using a higher-repetition-rate laser source.

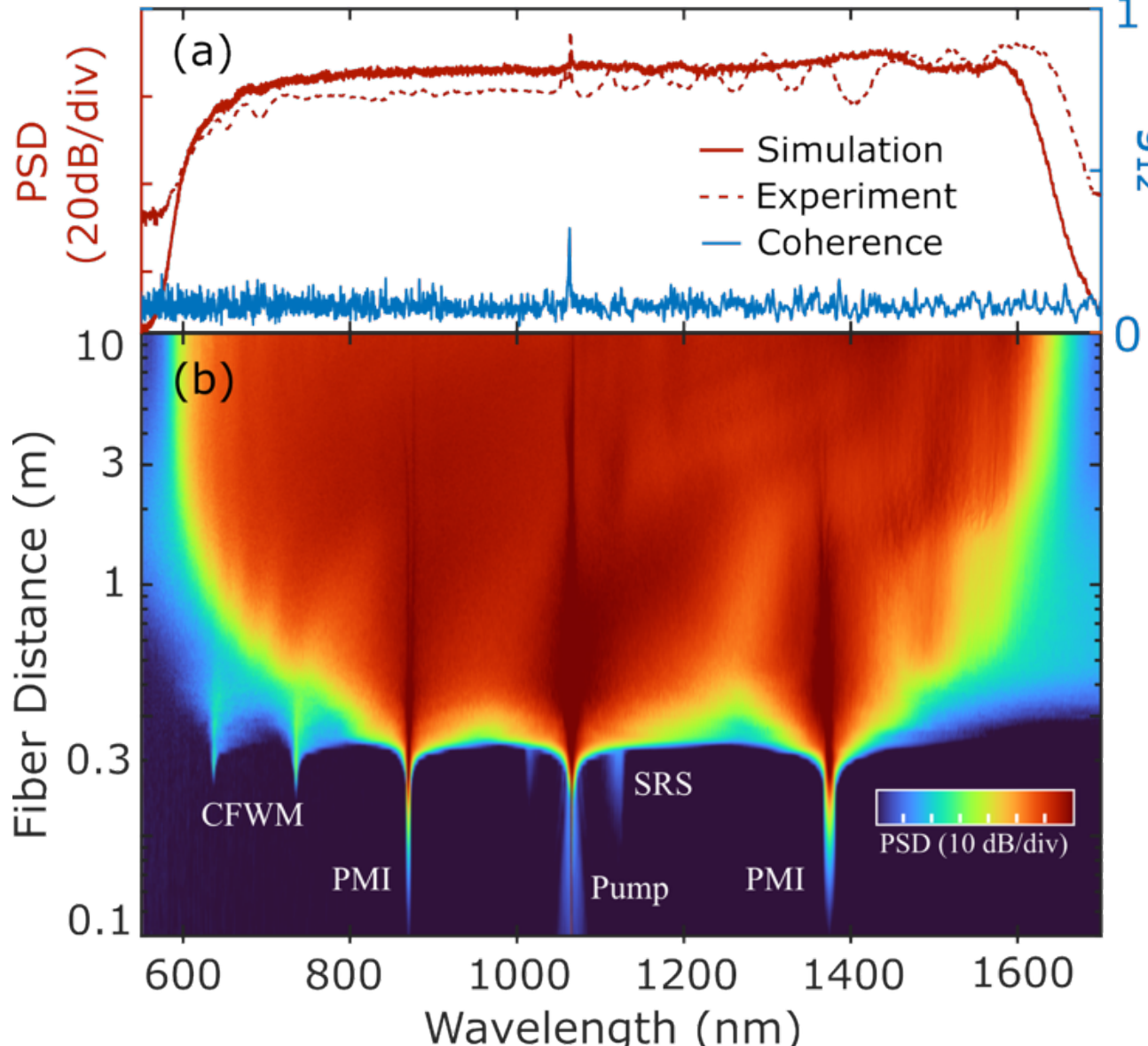

Fig. 5. (a) Experimental (red dashed) and numerical (red) SC spectra generated in the PM-ANDi-PCF when pumped at 1064 nm along the slow axis. The blue curve shows the SC spectral coherence $g_{12}$ (b) Numerically calculated SC spectral evolution (heatmap) as a function of propagation distance, illustrating SC generation triggered by PMI and SRS. Parameters are P=10 kW; L = 10 m.

Note also that the fiber length required to generate the SC in this experiment is roughly an order of magnitude longer than in conventional femtosecond SC regimes, which typically require only a few tens of centimeters of fiber [9,10]. In contrast, the SC reported here is governed primarily by PMI and SRS as the dominant broadening mechanisms, and the resulting bandwidth is not determined solely by the input peak power. This regime enables the generation of a comparable spectral bandwidth while using approximately half the peak power, albeit at the expense of a significantly longer propagation length.

In conclusion, we have experimentally demonstrated SC generation spanning more than one octave in the all-normal dispersion regime using long picosecond laser pulses. We have shown that this SC is enabled by the interplay between polarization modulation instability (PMI) and stimulated Raman scattering (SRS). Our results expand the potential applications of polarization-maintaining all-normal dispersion (PM-ANDi) fibers and highlight the constructive role that polarization instability can play, contrary to its conventional perception as a detrimental effect in standard femtosecond SC generation schemes. Notably, to the best of our knowledge, this is the first observation of PMI in high-birefringence fibers exhibiting such a large frequency shift. Previous reports have primarily documented PMI in low-birefringence fibers, making our results particularly significant for the understanding of vector nonlinear dynamics in strongly birefringent systems.

**Funding.** This work was supported by the EU Horizon program (grant No. 101135904, VISUAL), ANR Région Bourgogne Franche-Comté, and the Institut Universitaire de France. PhLAM authors also acknowledge CPER WaveTech and associated regional and national funding bodies.

**Data Availability Statement (DAS).** The data that support the findings of this study are available from the corresponding author upon reasonable request.

**Disclosures.** The authors declare no conflicts of interest.

**Full References**